# An all-fiber laser oscillating directly at single $TE_{01}$ mode through ring-core fibers

Yimin Zhang,[1] Hongxun Li,[1,2] Chuansheng Dai,[1] Runxia Tao,[1] Lixin Xu,[1] Chun Gu,[1] Wei Chen,[3] Yonggang Zhu,[3] Peijun Yao[1,*], and Qiwen Zhan[4,5]

[1]*Department of Optics and Optical Engineering, University of Science and Technology of China, Hefei, Anhui 230026, China*

[2] *Research Center of Laser Fusion, China Academy of Engineering Physics, Mianyang, Sichuan 621900, China*

[3]*Jiangsu Hengtong Optical Fiber Technology Com.Ltd, Suzhou, Jiangsu 215000, China*

[4]*Department of Electro-Optics and Photonics, University of Dayton, Dayton, Ohio 45469, USA*

[5]*School of Optical-Electrical and Computer Engineering, University of Shanghai for Science and Technology, Shanghai 200093, China*

*yap@ustc.edu.cn

**ABSTRACT:** Cylindrical vector beams (CVBs) have a wide range of applications owing to their particular polarization characteristics and optical field distributions. For the first time, an azimuthally polarized fiber laser without any polarization controller is proposed and demonstrated experimentally. The scheme is based on a self-designed ring-core fiber and transverse mode filter (TMF). The ring-core fiber can break the degeneracy of $LP_{11}$ modes and make the $TE_{01}$ mode propagate stably in the fiber laser. The TMF, which made from the ring-core fiber by depositing a layer of Aluminum on the cladding surface, can effectively suppress the modes other than $TE_{01}$ mode. The fiber laser can stably operate at $TE_{01}$ mode with a narrow 30dB linewidth of 0.18nm, which indicates the laser is polarization-maintained. This study opens a new avenue toward the true application of CVBs fiber lasers.

Recently, researchers have shown an increased interest in cylindrical vector beams (CVBs) because of their unique axisymmetric polarization and doughnut-shaped optical field distributions [1]. These particular characteristics are highly desirable in a wide range of applications, including material processing [2], surface plasmon excitation [3, 4], optical trapping and manipulation [5], electron acceleration [6], high-resolution metrology [7, 8], optical storage [9] and mode division multiplexing systems [10-12]. The systems of generating CVBs in free space is rather complicated and expensive, which made researchers try to find other simpler and more flexible ways. Compared with free space devices, fiber lasers show the advantages of low cost, compactness, high efficiency, flexibility and excellent heat management. The second order mode ($LP_{11}$ mode) in few-mode fibers is a linear superposition of four degenerate vector modes, including $TM_{01}$ (radially polarized), $TE_{01}$ (azimuthally polarized), $HE_{21}$ (even) and $HE_{21}$ (odd) modes. In 2002, Grosjean *et al.* [13] first generated CVBs in fiber lasers by introducing an offset between a single-mode fiber (SMF) and a few-mode fiber (FMF). Since then, several kinds of methods for generating CVBs in fiber lasers have been proposed and demonstrated, such as lateral offset splicing technique [14-16], mode selective coupler (MSC) [17-19] and long-period fiber grating (LPFG) [20,21]. However, these methods mentioned above are all based on the conversion between $LP_{01}$ mode and $LP_{11}$ mode. In 2016, Liu *et al.* [22] demonstrated a fiber laser oscillating at $LP_{11}$ mode by employing a pair of few-mode fiber Bragg gratings (FMFBGs). Wang *et al.* [23] also reported a fiber laser oscillating at high-order mode by using a wavelength-division-multiplexing (WDM) mode selective coupler (MSC). Our group also demonstrated several fiber lasers operating at $LP_{11}$ mode based on the polarization dependence of FMFBGs [24] and metal-clad transverse mode filter [25]. However, the main drawback of the research up to now is that the $LP_{11}$ mode is highly degenerate in the common step-index fiber, and the $TE_{01}$, $TM_{01}$, $HE_{21}$ (even) and $HE_{21}$ (odd) modes would experience a strong coupling with each other. Thus, the polarization controller (PC) is indispensable to obtain a specific vector mode in fiber lasers, which is rather unstable and impractical.

In this letter, we propose and experimentally demonstrate an all-fiber laser that works on the single $TE_{01}$ mode and directly output highly stable $TE_{01}$ mode without any polarization controller. To the best of our knowledge, this is the first report on the single $TE_{01}$ mode oscillation in fiber lasers. A pair of matched FMFBGs, which are inscribed on the ring-core fiber, is used as cavity reflector and output coupler. The laser operates at a wavelength of 1064.9 nm with a narrow 30 dB spectrum width of 0.18 nm. One significant advantage of the fiber laser is its polarization maintaining due to it operates at the single $TE_{01}$ mode. More generally, we believe that our results will promote the CVBs fiber lasers to be truly applied in a wide range of applications.

In conventional fibers, CVB states are highly unstable because of the extremely small differences ($\sim 10^{-5}$) of effective refractive index ($n_{eff}$) within the $LP_{11}$ mode group. Thus, the four CVB modes [i.e. $TE_{01}$, $TM_{01}$, $HE_{21}$ (even) and $HE_{21}$ (odd) modes] would randomly couple to each other during propagation due to fiber bends, geometry imperfections and stress-induced perturbations [26]. While in ring-core fibers, the differences of effective refractive index within the $LP_{11}$ mode can reach up to $1 \times 10^{-4}$, which was experimentally confirmed a reasonable proxy for mode stability [27-29]. Figure 1(a) shows the measured refractive index profile of the fabricated ring-core passive fiber and ring-core ytterbium-doped fiber (RC-YDF), which have the same refractive index profile. Figure 1(b) shows the $n_{eff}$ of $HE_{11}$, $TE_{01}$, $TM_{01}$ and $HE_{21}$ (even and odd) modes as a function of wavelength, calculated by the full-vector finite element method (FEM). The minimum difference of $n_{eff}$ between these modes is $1.0 \times 10^{-4}$, which is big enough to break the modal degeneracy and decrease inter-mode coupling to a very low level.

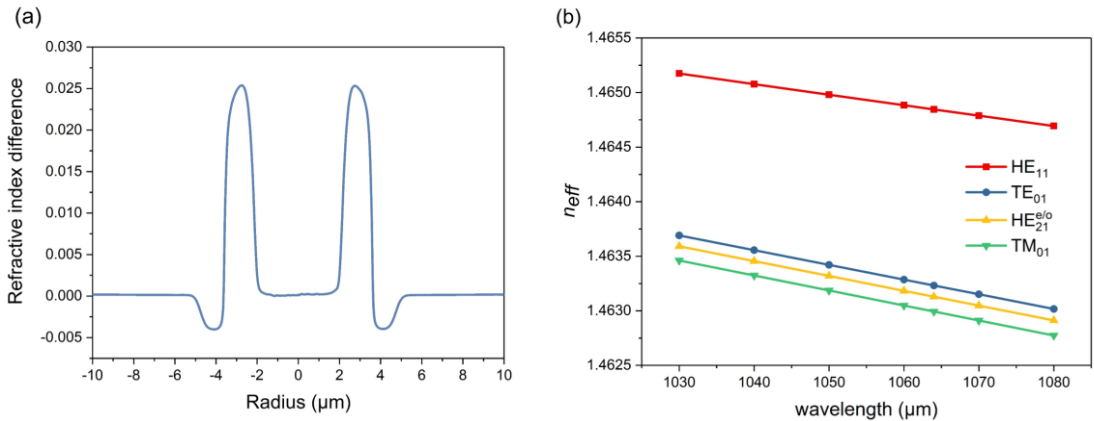

Fig. 1. (a) Refractive index profile of the fabricated ring-core fiber, (b) calculated $n_{eff}$ of $HE_{11}$, $TE_{01}$, $TM_{01}$ and $HE_{21}$ (even and odd) modes in ring-core fiber as the function of wavelength.

The schematic structure diagram of the transverse mode filter is shown in Fig. 2. It is made from the ring-core fiber by corroding the fiber cladding and depositing an Aluminum coating on the remained cladding surface. Figure 3 shows the scanning electron microscopy (SEM) images of the transverse mode filter which display a smooth surface and uniform diameter. The detailed fabrication process of the transverse mode filter was illustrated in the literature [25]. The transverse mode filter is almost lossless for $TE_{01}$ mode but not for other modes due to the different intensity distributions near the metal layer and cladding interface, the mechanism of which was explained in our previous work [25].

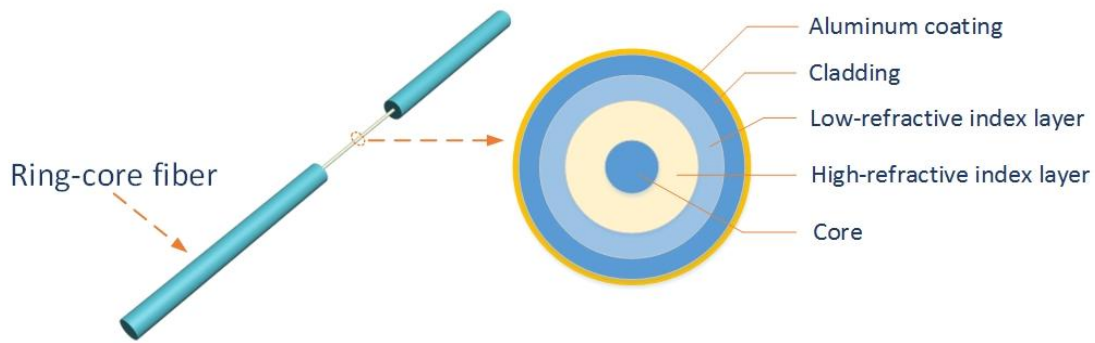

Fig. 2. Schematic of the transverse mode filter. The right part is the cross section of the transverse mode filter.

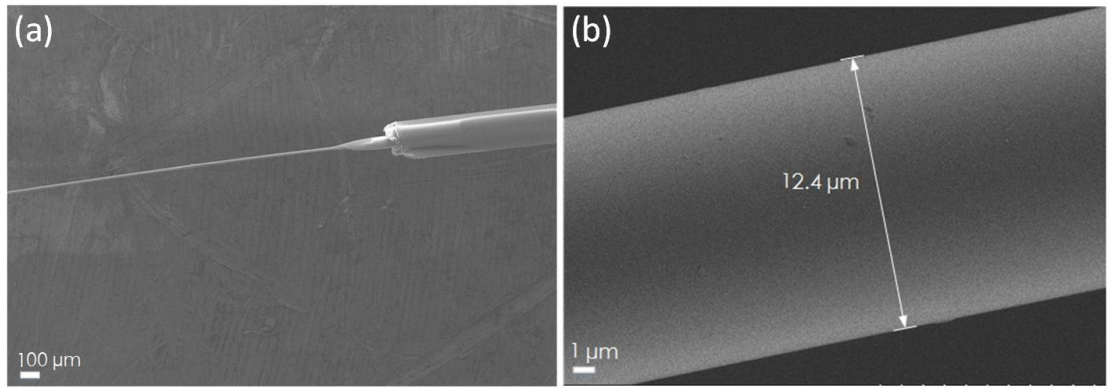

Fig. 3. The scanning electron microscopy images of the transverse mode filter at the magnification of (a) 45 and (b) 5000.

FEM is employed to calculate the $n_{eff}$ of the modes, Re($n_{eff}$) and Im($n_{eff}$) represent the real part and imaginary part of the effective refractive index, respectively. The confinement loss is given by [30]

$$Loss = \frac{20}{\ln 10} \frac{2\pi}{\lambda} \text{Im}(n_{eff}) \, dB/m \quad (1)$$

Figure 4 (a) and (c) show the Re($n_{eff}$) of modes $HE_{11}$, $TE_{01}$, $TM_{01}$ and $HE_{21}$ (even and odd) varying with the cladding diameter, where the thickness of the Aluminum coating is fixed at 200nm. It indicates that the difference of Re($n_{eff}$) between these modes would decrease as the cladding diameter decreases before the cladding diameter reaching a certain value (about 11.4 μm), thus the remained cladding of the transverse mode filter cannot be too thin in consideration of fabrication. In our fabrication, the remained cladding diameter is chosen to be around 12 μm. Figure 4 (b) and (d) show the confinement losses of the modes $HE_{11}$, $TE_{01}$, $TM_{01}$ and $HE_{21}$ (even and odd) varying with the thickness of Aluminum coating, where the cladding diameter is fixed at 12 μm. The results show that the confinement loss of $TE_{01}$ mode is as low as 0.03 dB/cm, while that of the other modes are large than 12 dB/cm. Thus, the transverse mode filter can make the $TE_{01}$ mode pass through while greatly attenuates the other modes. Considering the robustness and cost, the fabricated transverse mode filter is deposited an Aluminum layer of 200 nm.

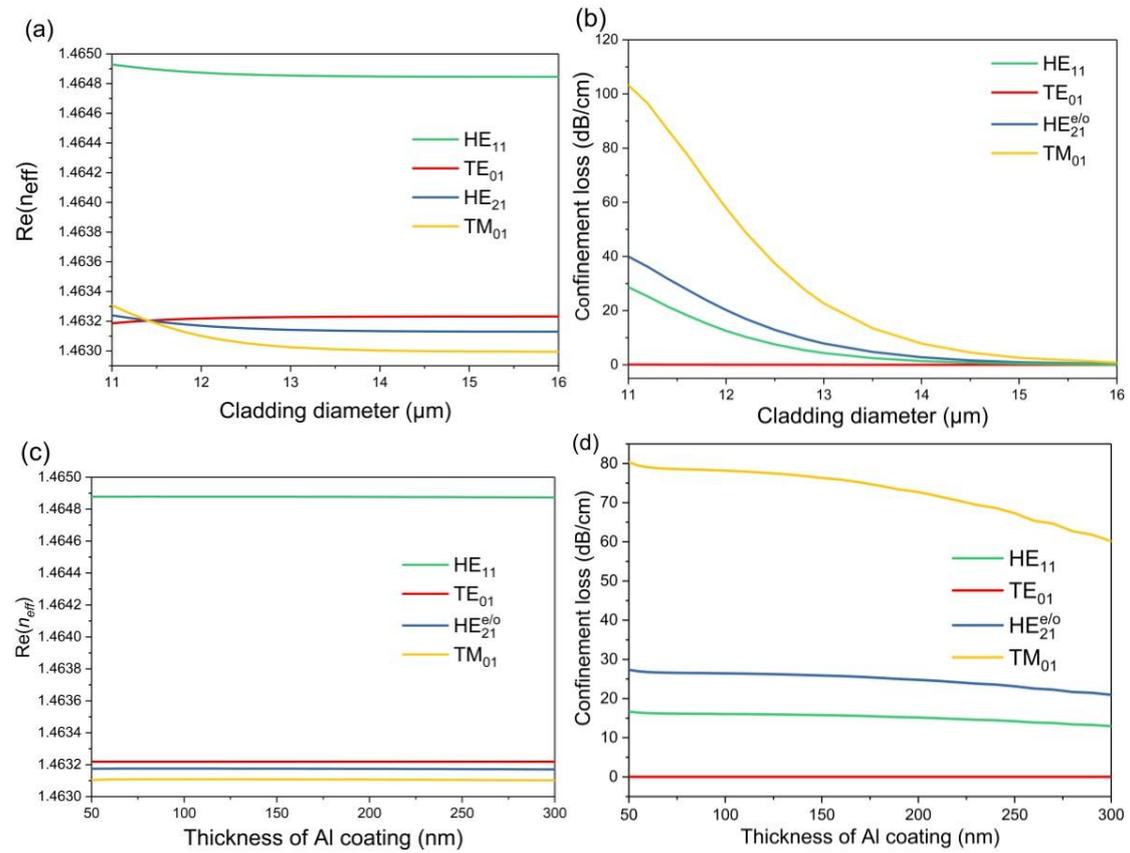

Fig. 4. Dependencies of Re($n_{eff}$) for the modes on the (a) cladding diameter and (c) thickness of Aluminum coating. Dependencies of confinement losses for the modes on the (b) cladding diameter and (d) thickness of Aluminum coating.

The schematic of azimuthally polarized fiber laser is shown in Fig. 5. A length of ~60 cm ring-core ytterbium-doped fiber is used as the gain medium and pumped by a 980nm laser diode through a 980/1064 nm wavelength division multiplexing. Two matched fiber Bragg gratings (FBGs), which are written on the ring-core passive fiber, are used as the feedback elements and output coupler. The transverse mode filter (TMF) is used to attenuate $HE_{11}$, $TM_{01}$ and $HE_{21}$ (even and odd) modes, and make $TE_{01}$ mode oscillate in the cavity. The output beams are recorded using a CCD camera after collimating by lenses $L_1$ and $L_2$.

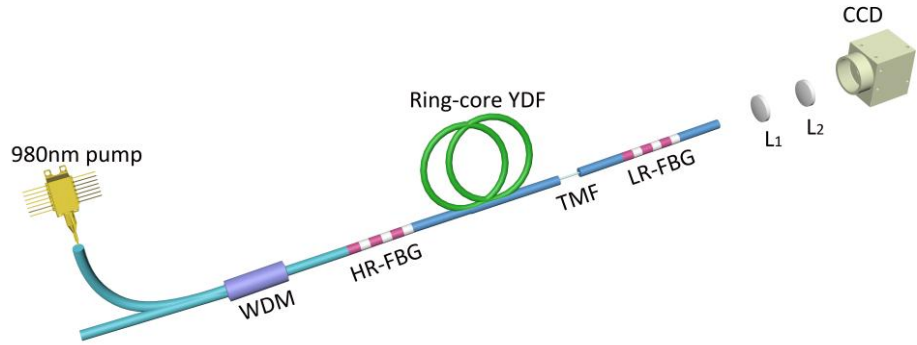

Fig. 5. The experiment setup of the azimuthally polarized fiber laser. WDM: 980/1064nm wavelength division multiplexing; HR-FBG: high reflective fiber Bragg grating; Ring-core YDF: ring-core ytterbium-doped fiber; TMF: transverse mode filter; LR-FBG: low reflective fiber Bragg grating; $L_1$, $L_2$: biconvex lens; CCD: charge coupled device, infrared camera.

The transmission spectra of the FBGs are measured using a 1.0 μm amplified spontaneous emission (ASE) light source when only $LP_{01}$ mode is excited, and the peak reflectivity is estimated to be 93% (~11 dB) for HR-FBG and about 75% (~6 dB) for LR-FBG. In theory, the reflectivity of $LP_{11}$ mode is close to $LP_{01}$ mode. The reflection spectrum of the FBG, shown in Fig. 6 as the blue curve, has four reflection peaks (marked as 1, 2, 3, and 4) which correspond to the intra-modal reflection peaks of $TM_{01}$, $HE_{21}$ (even and odd), $TE_{01}$ and $HE_{11}$ modes, respectively. The laser spectrum is measured by the optical spectrum analyzer (OSA) at the output port of LR-FBG, as shown in Fig. 6 as the orange curve. The laser operates at a single wavelength of 1064.8 nm with a narrow 30 dB linewidth of 0.18 nm and side-mode suppression ratio (SMSR) of more than 65 dB. Most importantly, it can be seen that the spectral peak is consistent with the intra-modal reflection peak of $TE_{01}$ mode, which indicates the laser is oscillating at the single $TE_{01}$ mode. As shown in Fig. 7, the slope efficiency of the laser is about 16.8% with a laser threshold of ~125 mW. The relatively low slope efficiency and high threshold are owing to the slight ellipticity of the fiber and the corrosion unevenness of the fabricated TMF, which cause the $TE_{01}$ mode experience a higher loss when passing through the TMF.

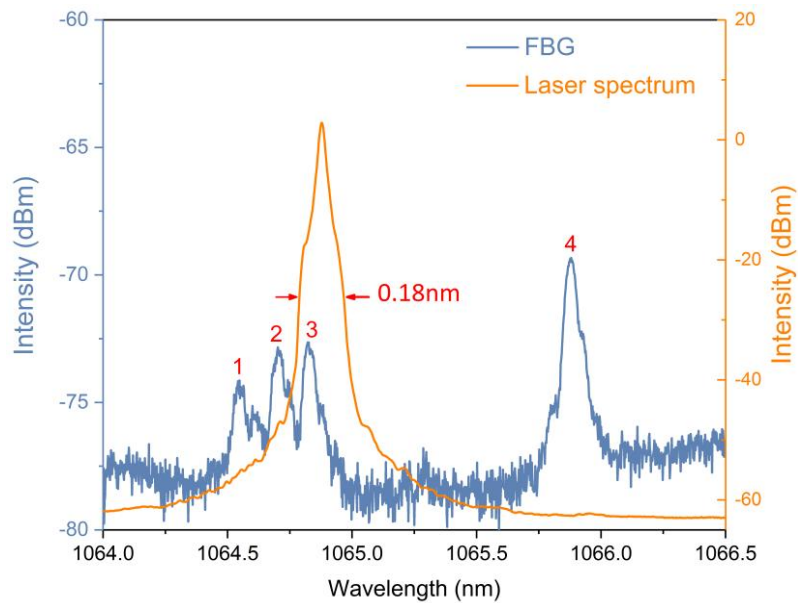

Fig. 6. Reflection spectrum of the FBG (blue line). Output spectrum of the fiber laser (orange line).

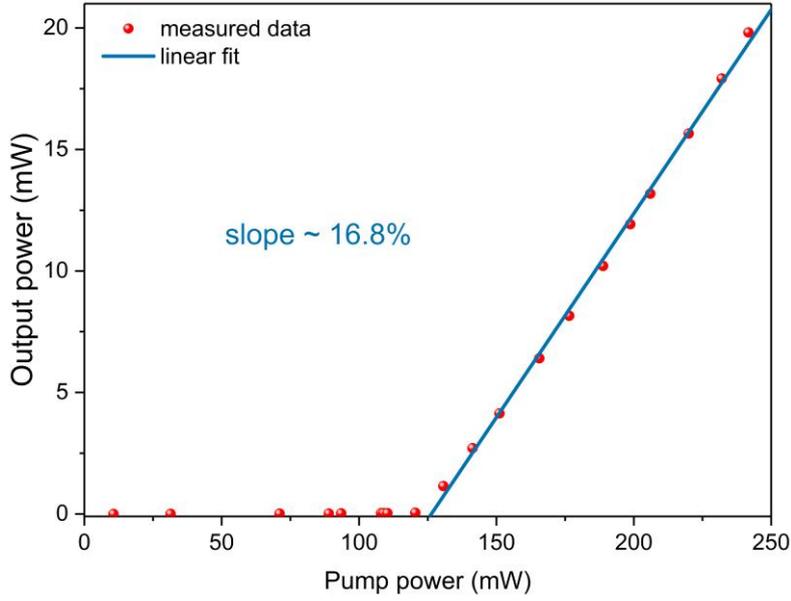

Fig. 7. Output power characteristic of the fiber laser.

Figure 8 shows the intensity distributions of output beam recorded by CCD camera. The typical doughnut-shaped intensity distribution is shown in Fig. 8(a), and its polarization property is examined by placing a linear polarizer between lenses $L_1$ and $L_2$, as shown in Fig. 8(b)-8(e), proving the azimuthal polarization distribution. The polarization degree of output beam is measured using an improved method based on the literature [31]. For an ideal azimuthally polarized vector beam, the light is linearly polarized along the azimuthal direction at any point of the ring profile. Thus, the polarization degree of the vector beam can be represented by the polarization extinction ratio (PER) of a localized point. Based on this principle, an aperture with 1 mm diameter is placed on the propagation path of output beam behind output coupler, where the diameter of the beam profile is measured to be about 7 cm. Due to the size of the aperture is far smaller than the beam profile, the light passing through the aperture can be considered to be approximately linearly polarized. Consequently, the PER of passed light can be measured by rotating a polarizer. The PER is calculated by

$$PER = \frac{P_{azimuthal} - P_{radial}}{P_{azimuthal} + P_{radial}} \qquad (2)$$

Where $P_{azimuthal}$ and $P_{radial}$ are the power of passed light when the direction of the polarizer is azimuthal and radial, respectively. In order to reduce the measurement error, the PER of eight different points at the beam profile is measured, indicated with the white crosses as shown in Fig. 9, and the average polarization degree is 94.4%.

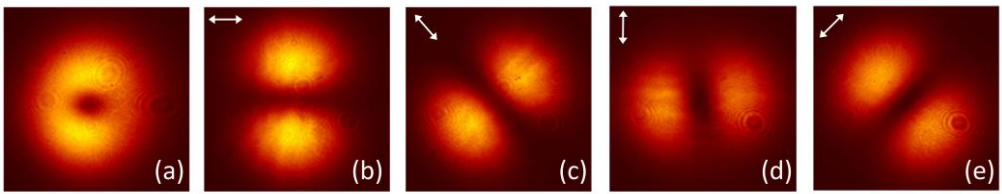

Fig. 8. (a) Intensity distribution of the output beam; (b)-(e) intensity distributions after passing through a linear polarizer. The white arrows indicate the transmission axis orientations of the polarizer.

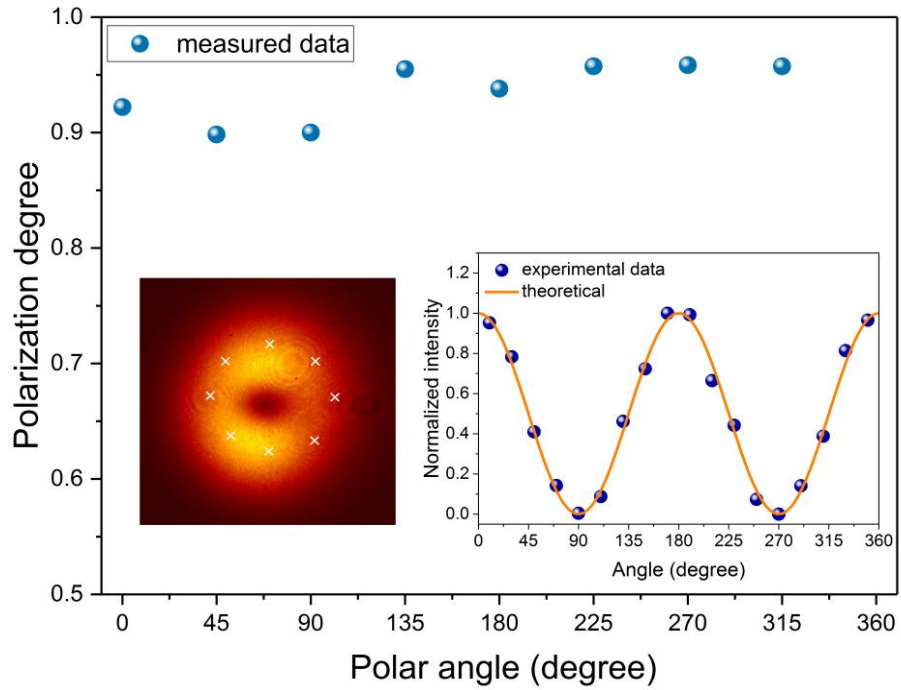

Fig. 9. Polarization degrees at different positions of the beam profile (blue dot), and the white crosses inside the left inset represent the position of aperture. The inset on the right is the intensity variation of the transmitted light when rotating the polarizer 360 degrees.

To estimate the mode purity of output beam, one-dimensional intensity distribution analysis proposed in the literature [32] is used. The horizontal and the vertical one-dimensional intensity distributions of output beam are measured as shown in Fig. 10, which are fitted by the linear superposition of $LP_{01}$ and $TE_{01}$ modes. The powers of different modes are obtained by calculating the area under the curves, and the average mode purity of $TE_{01}$ is calculated to be 97.5%.

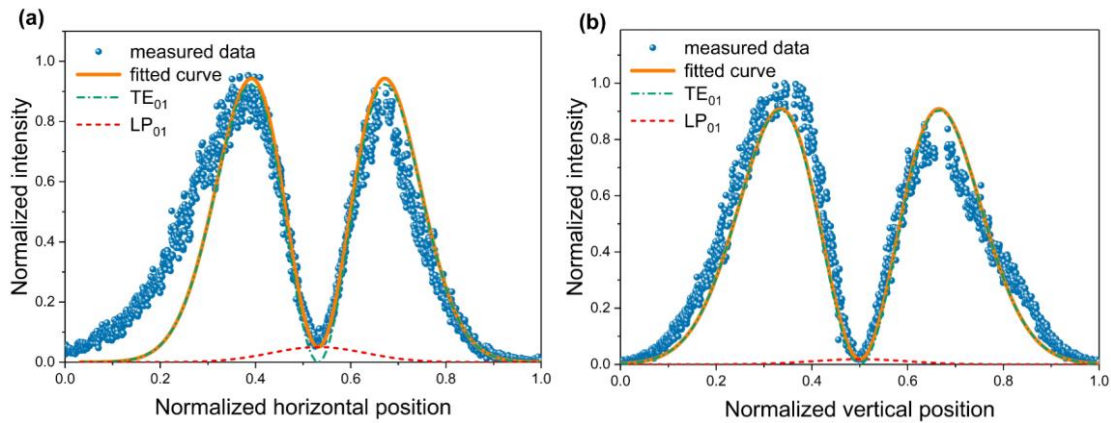

Fig. 10. (a) Horizontal and (b) vertical intensity distribution across the center of the output beam. The blue dots represent the measured intensity profiles, and the orange curves are the linear superposition of $LP_{01}$ and $TE_{01}$ modes. The green and the red dashed curves correspond to the theoretical one-dimensional intensity distribution of $TE_{01}$ and $LP_{01}$ modes, respectively.

In conclusion, we have proposed and demonstrated experimentally an azimuthally polarized fiber laser. By using ring-core fibers to break the degeneracy of LP$_{11}$ modes and a transverse mode filter to suppress the modes other than TE$_{01}$ mode, the laser can stably work on the single TE$_{01}$ mode with a narrow 30 dB linewidth of 0.18 nm at the center wavelength of 1064.8 nm. More importantly, operating at a single TE$_{01}$ mode means the fiber laser is polarization-maintained. We believe our results will give a strong impetus to the applications of cylindrical vector beam fiber lasers in areas such as material processing, surface plasmon excitation, optical tweezers, electron acceleration, high-resolution metrology, optical storage and mode division multiplexing systems.

**Funding**

This work was supported by the National Natural Science Foundation of China (No.61675188); the Fundamental Research Funds for the Central Universities (WK6030000065); Science and Technology on Plasma Physics Laboratory (No.6142A0403060917).

**Acknowledgments**

This work was partially carried out at the USTC Center for Micro and Nanoscale Research and Fabrication, and we thank Yu Wei for his help on micro fabrication.

**References**

1. Q. Zhan, "Cylindrical vector beams: from mathematical concepts to applications," Adv. Opt. Photonics **1(1)**, 1-57 (2009).
2. M. Meier, V. Romano, and T. Feurer, "Material processing with pulsed radially and azimuthally polarized laser radiation," Appl. Phys. A Mater. Sci. Process. **86(3)**, 329–334 (2007).
3. B. N. Tugchin, N. Janunts, A. E. Klein, M. Steinert, S. Fasold, S. Diziain, M. Sison, E. B. Kley, A. Tünnermann, and T. Pertsch, "Plasmonic Tip Based on Excitation of Radially Polarized Conical Surface Plasmon Polariton for Detecting Longitudinal and Transversal Fields," ACS Photonics **2(10)**, 1468–1475 (2015).
4. M. Piliarik, J. Homola, Z. Manıková, and J. Čtyroký. "Surface plasmon resonance sensor based on a single-mode polarization-maintaining optical fiber." Sensors and Actuators B: Chemical **90(1-3)**, 236-242 (2003).
5. O. M. Maragò, P. H. Jones, P. G. Gucciardi, G. Volpe, and A. C. Ferrari, "Optical trapping and manipulation of nanostructures," Nat. Nanotechnol. **8(11)**, 807–819 (2013).
6. S. R. Greig and A. Y. Elezzabi, "Generation of attosecond electron packets via conical surface plasmon electron acceleration," Sci. Rep. **6**, 1–9 (2016).
7. G. Bautista, C. Dreser, X. Zang, D. P. Kern, M. Kauranen, and M. Fleischer, "Collective Effects in Second-Harmonic Generation from Plasmonic Oligomers," Nano Lett. **18(4)**, 2571–2580 (2018).
8. G. Bautista, J.-P. Kakko, V. Dhaka, X. Zang, L. Karvonen, H. Jiang, E. Kauppinen, H. Lipsanen, and M. Kauranen, "Nonlinear microscopy using cylindrical vector beams: Applications to three-dimensional imaging of nanostructures," Opt. Express **25(11)**, 12463–12468 (2017).
9. M. Gu, X. Li, and Y. Cao, "Optical storage arrays: A perspective for future big data storage," Light Sci. Appl. **3(5)**, e177 (2014).

10. N. Bozinovic, Y. Yue, Y. Ren, M. Tur, P. Kristensen, H. Huang, A. E. Willner, and S. Ramachandran, "Terabit-scale orbital angular momentum mode division multiplexing in fibers," Science **340 (6140)**, 1545–1548 (2013).
11. D. J. Richardson, J. M. Fini, and L. E. Nelson, "Space-division multiplexing in optical fibres," Nat. Photonics **7(5)**, 354–362 (2013).
12. J. Liu, S. Li, L. Zhu, A. Wang, S, C. Klitis, C. Du, Q. Mo, M. Sorel, S. Yu, X. Cai, and J. Wang, "Direct fiber vector eigenmode multiplexing transmission seeded by integrated optical vortex emitters," Light Sci. Appl. **7(3)**, 17148 (2018).
13. T. Grosjean, D. Courjon, and M. Spajer, "An all-fiber device for generating radially and other polarized light beams," Opt. Commun. **203(1-2)**, 1–5 (2002).
14. B. Sun, A. Wang, L. Xu, C. Gu, Z. Lin, H. Ming, and Q. Zhan, "Low-threshold single-wavelength all-fiber laser generating cylindrical vector beams using a few-mode fiber Bragg grating.," Opt. Lett. **37(4)**, 464–466 (2012).
15. J. Lin, K. Yan, Y. Zhou, L. X. Xu, C. Gu, and Q. W. Zhan, "Tungsten disulphide based all fiber Q-switching cylindrical-vector beam generation," Appl. Phys. Lett. **107(19)**, 191108 (2015).
16. K. Yan, J. Lin, Y. Zhou, C. Gu, L. Xu, A. Wang, P. Yao, and Q. Zhan, "Bi_2Te_3 based passively Q-switched fiber laser with cylindrical vector beam emission," Appl. Opt. **55(11)**, 3026-3029 (2016).
17. Z. Zhang, Y. Cai, J. Wang, H. Wan, and L. Zhang, "Switchable dual-wavelength cylindrical vector beam generation from a passively mode-locked fiber laser based on carbon nanotubes," IEEE J. Sel. Top. Quantum Electron. **24(3)**, 1–6 (2018).
18. H. O. W. An, J. I. E. W. Ang, Z. U. Z. Hang, Y. U. C. Ai, and B. I. N. S. Un, "High efficiency mode-locked , cylindrical vector beam fiber laser based on a mode selective coupler," Opt. Express **25(10)**, 1–8 (2017).
19. J. Wang, H. Wan, H. Cao, Y. Cai, B. Sun, Z. Zhang, and L. Zhang, "A 1.0 μm Cylindrical Vector Beam Fiber Ring Laser Based on A Mode Selective Coupler," IEEE Photonics Technol. Lett. **30(9)**, 765-768 (2018).
20. Y. Zhou, K. Yan, R. S. Chen, C. Gu, L. X. Xu, A. T. Wang, and Q. Zhan, "Resonance efficiency enhancement for cylindrical vector fiber laser with optically induced long period grating," Appl. Phys. Lett. **110(16)**, 161104 (2017).
21. R. Chen, J. Wang, X. Zhang, A. Wang, H. Ming, F. Li, D. Chung, and Q. Zhan, "High efficiency all-fiber cylindrical vector beam laser using a long-period fiber grating," Opt. Lett. **43(4)**, 755-758 (2018).
22. T. Liu, S. Chen, and J. Hou, "Selective transverse mode operation of an all-fiber laser with a mode-selective fiber Bragg grating pair," Opt. Lett. **41(24)**, 5692-5695 (2016).
23. T. Wang, F. Shi, Y. Huang, J. Wen, Z. Luo, F. Pang, T. Wang, and X. Zeng, "High-order mode direct oscillation of few-mode fiber laser for high-quality cylindrical vector beams," Opt. Express **26(9)**, 11850-11858 (2018).
24. H. Li, K. Yan, Y. Zhang, C. Gu, P. Yao, L. Xu, R. Zhang and J. Su，" Low-threshold high-efficiency all-fiber laser generating cylindrical vector beams operated in $LP_{11}$ mode throughout the entire cavity." Appl. Phys. Express **11**. 122502 (2018).
25. Y. Zhang, H. Li, C. Dai, L. Xu, C. Gu, W. Chen, Y. Zhu, P. Yao and Q. Zhan, "An all-fiber high-order mode laser using a surface plasmonic transverse mode filter" Opt. Express **26(23)**. 29679-29686 (2018).

26. S. Ramachandran and P. Kristensen, "Optical vortices in fiber," Nanophotonics **2(5-6)**, 455–474 (2013).
27. S. Ramachandran, P. Kristensen, and M. F. Yan, "Generation and propagation of radially polarized beams in optical fibers.," Opt. Lett. **34(16)**, 2525–2527 (2009).
28. P. Gregg, P. Kristensen, and S. Ramachandran, "Conservation of orbital angular momentum in air-core optical fibers: erratum," Optica **2(3)**, 267-270 (2017).
29. S. Ramachandran, P. Gregg, P. Kristensen, and S. E. Golowich, "On the scalability of ring fiber designs for OAM multiplexing," Opt. Express **23(3)**, 3721-3730 (2015).
30. J. Li, C. Wang, and W. Wang, "Generation of an azimuthally polarized beam with a metallic ring core fiber.," Appl. Opt. **52(32)**, 7759–7768 (2013).
31. D. Lin, K. Xia, J. Li, R. Li, K. Ueda, G. Li, and X. Li, "Efficient, high-power, and radially polarized fiber laser," Opt. Lett. **35(13)**, 2290–2292 (2010).
32. D. Lin, J. M. O. Daniel, M. Gecevičius, M. Beresna, P. G. Kazansky, and W. A. Clarkson, "Cladding-pumped ytterbium-doped fiber laser with radially polarized output," Opt. Lett. **39(18)**, 5359-5361 (2014).